\documentclass[useAMS,usenatbib]{mn2e}
\usepackage{epsfig,bm,subfigure}
\usepackage{amsmath, amssymb}

\title[Alignment Timescale of the Microquasar GRO J1655-40]{Alignment
  Timescale of the Microquasar GRO J1655-40}

\author[R. G. Martin, C. A. Tout \& J. E. Pringle]{Rebecca G. Martin,
  Christopher A. Tout and J. E. Pringle\\University of Cambridge,
  Institute of Astronomy, The Observatories, Madingley Road, Cambridge
  CB3 0HA\\}
\begin{document}
\date{}

\pagerange{\pageref{firstpage}--\pageref{lastpage}} \pubyear{2007}
\maketitle

\label{firstpage}

\begin{abstract}
  The microquasar GRO~J1655--40 has a black hole with spin angular
  momentum apparently misaligned to the orbital plane of its companion
  star.  We analytically model the system with a steady state disc
  warped by Lense-Thirring precession and find the timescale for the
  alignment of the black hole with the binary orbit.  We make detailed
  stellar evolution models so as to estimate the accretion rate and
  the lifetime of the system in this state.  The secondary can be
  evolving at the end of the main sequence or across the Hertzsprung
  gap. The mass-transfer rate is typically fifty times higher in the
  latter case but we find that, in both cases, the lifetime of the mass
  transfer state is at most a few times the alignment timescale.  The
  fact that the black hole has not yet aligned with the orbital plane
  is therefore consistent with either model. We conclude that the
  system may or may not have been counter-aligned after its supernova
  kick but that it is most likely to be close to alignment rather than
  counteralignment now.

\end{abstract}


\section{Introduction}

Microquasars are X-ray binaries which have relativistic radio jets
\citep{MR99}. They consist of a compact object such as a black hole or
a neutron star which is accreting matter from a companion star.  GRO
J1655--40 is one of only fifteen observed microquasars in our galaxy
\citep{P05} and has a black hole of mass $M_1=6.3\pm 0.5\,\rm M_\odot$
and a companion star of mass $M_2=2.4\pm 0.4\,\rm M_\odot$
\citep{G01}.

GRO J1655--40 shows transient behaviour in a complex form. It was
discovered in July 1994 \citep{zhang1994} through observations of hard
X-ray outbursts and then observed again several times in the following
year. Before 1994 it appeared to be quiescent for some time.  The
original outbursts were associated with radio flaring and superluminal
motion of the radio plasmoids which led \cite{H95} to estimate the
distance to the system to be $3.2\,\rm kpc$.

\cite{I99} measured the abundances of oxygen, magnesium, silicon and
sulphur in the atmosphere of the companion star to be six times
greater than in the Sun.  The secondary star does not have enough mass
to reach the internal temperatures required to create these elements.
A stellar wind or mass transfer from the primary during its
presupernova evolution could only have provided a small fraction of
these observed overabundances because the CNO cycle would have led to
a much larger nitrogen to oxygen ratio than is observed.  So these
abundances are interpreted as evidence that supernova ejecta have been
captured by the secondary star, perhaps during fall-back on to the
black hole. The relative abundances of the contaminating elements led
\cite{I99} to suggest that the black hole's progenitor was a star with
mass $25-40\,\rm M_\odot$.  It is therefore likely that the system
lost more than half its mass in the supernova even if the progenitor
of the black hole had had a strong stellar wind.  In order for the
system to have remained bound the supernova explosion must have been
associated with a substantial kick.  Such a kick could have altered
the spin of the black hole leaving it misaligned with the binary
orbit. Even a small velocity kick can lead to a large offset between
the spin of the black hole and the orbital axes \citep{BP95}.  There
is no reason why a kick substantial enough to keep the system bound
would leave the spin aligned with the orbit.

GRO J1655--40 has relativistic jets of material leaving the system
which we assume are generated in the inner parts of a black hole
accretion disc and so are perpendicular to the plane of the disc close
to the hole.  The combined action of the Lense-Thirring effect and the
internal viscosity of the accretion disc causes the angular momenta of
the black hole and the inner accretion disc to align. This is known as
the \cite{BP75} effect. It affects only the inner regions of the disc
because of the short range of the Lense-Thirring effect. The outer
parts of the disc tend to remain in their original configuration.

\cite{H95} measured the jet inclination to be $85^\circ \pm 2^\circ$
to the line of sight.  This differs significantly from the binary
orbital plane inclination of $70.2^\circ \pm 1.9^\circ$ \citep{G01}.
This implies that there is a misalignment between the inclination of
the black hole and the outer parts of the accretion disc and so the
accretion disc is warped.  The inner parts are aligned with the black
hole spin and the outer parts are aligned with the binary orbital
plane because of tidal torques.

\begin{figure}
\epsfxsize=8.4cm
\epsfbox{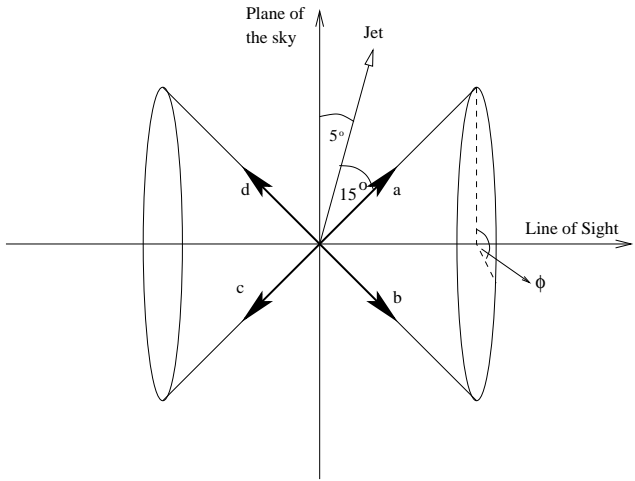}
\caption[]{We consider the position of the jet fixed relative to the line of sight. 
  The cone represents the surface on which the binary orbital angular
  momentum must lie.  The smallest possible inclination between the
  black hole and the binary angular momentum occurs in position
  {\bf{a}} where $\eta=15^\circ$.  The largest inclination between the
  black hole and the binary angular momentum occurs in position
  {\bf{c}} where $\eta=165^\circ$.}
\label{diagram}
\end{figure}

In Figure~1 we show the line of sight and the given position of the
jet. Relative to this jet, the angular momentum vector of the binary
must lie in one of the cones either pointing towards or away from us.
The smallest possible inclination between the black hole and the
binary angular momentum, $\eta$, occurs in position {\bf{a}} where
$\eta=15^\circ$ and the system is close to alignment.  The largest
possible inclination between the black hole and the binary angular
momentum occurs in position {\bf{c}} where $\eta=165^\circ$ and the
system is close to counter-alignment.  The inclination of the system
is in the range $15^\circ<\eta<165^\circ$.

\cite{King05} find that for a black hole that is massive compared to a
misaligned disc, then the disc aligns with the black hole. If the disc
is massive compared with the black hole then the black hole aligns
with the disc. In GRO J1655--40 the disc is much less massive than the
black hole. However because the disc is in, and is tidally linked to
the orbitial motion of, a binary system, what matters is the amount of
angular momentum in the binary orbit, not just in the disc. The
orbital angular momentum is several hundred times that of the black
hole so we expect the hole to tend to eventually align with the
orbit.

\cite{Macc} uses the model of \cite{NP98} to calculate the timescale
on which the black hole aligns with the outer disc and hence the
binary orbit. This assumes that the surface density and viscosities
are constant.  He finds that the ratio of alignment timescale to
binary lifetime is about $0.3$. He assumes the binary lifetime is the
time to accrete the whole star, not the stellar evolution time.  Using
the outburst history of GRO J1655--40 \cite{Macc} finds the alignment
timescale to be $8\times 10^8\,\rm yr$ with a companion of
main-sequence lifetime of $7\times 10^8\,\rm yr$.  He concludes that
the system should be approximately one alignment timescale in age and
so the black hole should not be fully aligned with the disc because it
takes more than one alignment timescale for full alignment.

We examine in more detail the stellar evolution of the companion star
to estimate how long the system can remain in this steady state and
whether the system should be aligned or not.  We make use of the
analysis of \cite{MPT07} and use power laws in distance from the
central black hole for the surface density and viscosities to find the
alignment timescale of the system. We consider the possibility that
the disc was initially counter-rotating.

\section{Misalignment}

Because the binary is not visually separated we only know the
inclination of the orbit relative to the line of sight. We do not know
its projection on to the sky. Given the observed inclination of the
binary orbit and the jets to the line of sight we can find the
probability distribution of the angle between the two momenta.

If the position angle, $\phi$, of the binary relative to the line of
sight (see Figure~\ref{diagram}) is randomly distributed in
$0<\phi<2\pi$ then
\begin{equation}
P(\phi) d\phi= \frac{1}{2\pi} d\phi.
\end{equation}
Since
\begin{equation}
P(\eta)d\eta=P(\phi) d\phi
\end{equation}
we deduce
\begin{equation}
P(\eta)=P(\phi)\frac{d\phi}{d\eta}=\frac{1}{2\pi} \frac{d\phi}{d\eta}.
\end{equation}
Note that each value of $\eta$ maps to two values of $\phi$.  If the
binary angular momentum is pointing towards us we have the standard
spherical trigonometric cosine formula
\begin{equation}
\cos \eta = \cos i \cos j + \sin i \sin j \cos \phi.
\end{equation}
where $i=85^\circ$ is the inclination of the jet to the line of sight
and $j=70^\circ$ is the inclination of the binary orbit to the line of
sight.  Differentiating we find
\begin{equation}
\frac{d\phi}{d \eta} = \frac{ \sin \eta}{\sin i \sin j \sin \phi}.
\end{equation}
If the binary orbital angular momentum is pointing away from us then
we have the relation
\begin{equation}
\cos \eta = \cos i \cos (\pi-j) + \sin i \sin (\pi-j) \cos \phi
\end{equation}
and we find
\begin{equation}
\frac{d\phi}{d \eta} = \frac{ \sin \eta}{\sin i \sin (\pi-j) \sin \phi}.
\end{equation}

\begin{figure}
  \includegraphics[width=8.4cm]{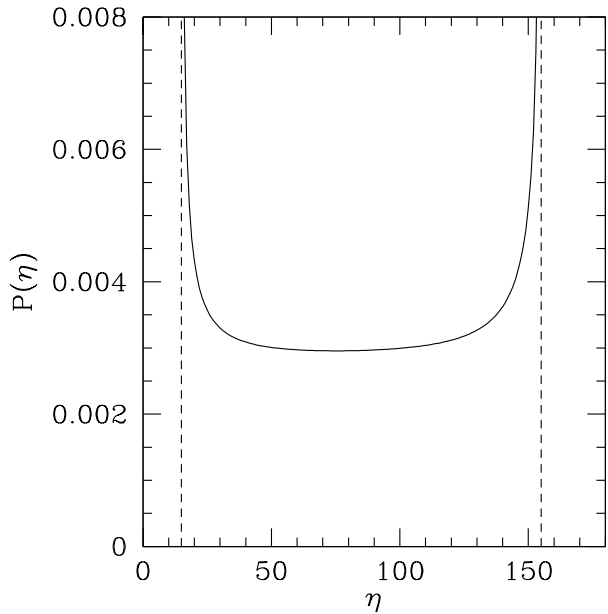}
\caption{The probability distribution of the angle between
  the jet and the binary orbit, $\eta$, when the angular momentum of
  the binary points towards us.}
\label{fig:prob1}
\end{figure}

\begin{figure}
  \includegraphics[width=8.4cm]{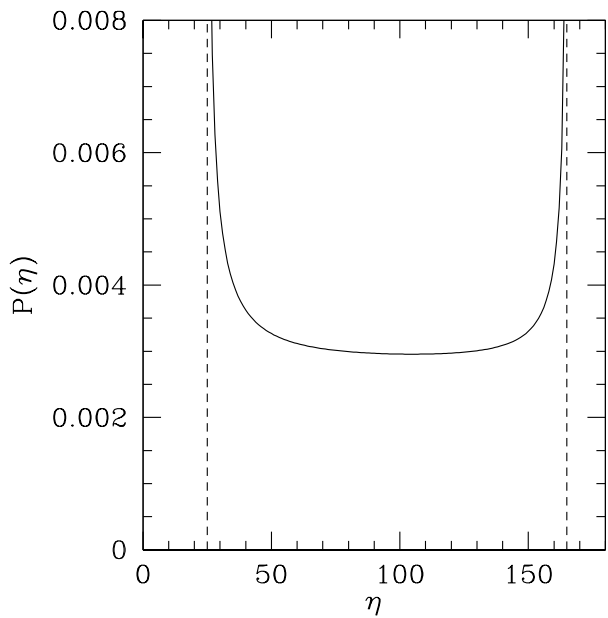}
\caption{The probability distribution of the angle between 
  the jet and the binary orbit, $\eta$, when the angular momentum of
  the binary points away from us.}
\label{fig:prob2}
\end{figure}

In Figures~\ref{fig:prob1} and~\ref{fig:prob2} we plot the probability
distributions for the two cases where the binary angular momentum is
pointing towards and away from us. We assume either is equally likely.
We see that if the binary points towards us, the most likely
misalignment angles are the extremes of $\eta=15^\circ$ (case {\bf a}
in Figure~1) or $155^\circ$ (case {\bf b}) and if the binary points
away then the most likely angles are $\eta=25^\circ$ (case {\bf d}) or
$165^\circ$ (case {\bf c}).

\section{Warped Accretion Discs}
\label{gen}

There are two viscosities in the disc, $\nu_1$ corresponds to the
azimuthal shear (the viscosity normally associated with accretion
discs) and $\nu_2$ corresponds to the vertical shear in the disc which
smoothes out the twist. The second viscosity acts when the disc is
non-planar.  We assume that we have a steady state disc in which
$\nu_1 \Sigma=\,\rm const$ and that the surface density is a power law
\begin{equation}
\Sigma=\Sigma_0\left(\frac{R}{R_0}\right)^{-\beta} 
\end{equation} 
\citep{SS73}, where $R$ is the spherical radial coordinate, $R_0$ is
some fixed radius and $\Sigma_0$ is a constant. To be in steady state,
the first viscosity must obey
\begin{equation} 
\nu_1=\nu_{10}\left(\frac{R}{R_0}\right)^{\beta},   
\label{visc1}
\end{equation}
where $\beta$ and $\nu_{10}$ are constants. We assume that
the second viscosity also obeys the power law
\begin{equation}
\nu_2=\nu_{20}\left(\frac{R}{R_0}\right)^{\beta}
\label{visc2}
\end{equation}
and $\nu_{20}$ is a constant.

Following the work of \cite{MPT07}, we consider a black hole of mass
$M_1$ at $R=0$ with spin angular momentum $\bm{J}$. We consider
the disc to be made up of annuli of width $dR$ and mass $2\pi \Sigma R
dR$ at radius $R$ from the central object of mass $M_1$ with
surface density $\Sigma(R,t)$ at time $t$ and with specific angular
momentum density $\bm{L}=(GM_1R)^{1/2}\Sigma \bm{l}=L\bm{l}$.
The unit vector describing the direction of the angular momentum of a
disc annulus is $\bm{l}=(l_x,l_y,l_z)$ with $|\bm{l}|=1$.  We
use equation (2.8) of \cite{P92} setting $\partial \bm{L}/ \partial t
= 0$ and add a term to describe the Lense-Thirring precession (the
last one) to give
\begin{align}
0=&\frac{1}{R}\frac{\partial}{\partial R}\left[ \left( \frac{3R}{L} \frac{\partial}{\partial R}(\nu_1 L)
  -\frac{3}{2}\nu_1\right)\bm{L}+\frac{1}{2}\nu_2RL\frac{\partial \bm{l}}{\partial R}\right] \cr
 & + \frac{\bm{\omega_{\rm p}} \times \bm{L}}{R^3}.
\label{main}
\end{align}
The Lense-Thirring precession is given by
\begin{equation}
\bm{\omega_{\rm p}} =\frac{2G\bm{J}}{c^2}
\label{omegap}
\end{equation}
\citep{KP85}, where the angular momentum of the black hole can be
expressed in terms of the dimensionless spin parameter $a$ such that
\begin{equation}
J=acM_1\left(\frac{GM_1}{c^2}\right).
\label{angmom}
\end{equation}
The black hole spin is also evolving because of the torques exerted by
the disc so we have
\begin{equation}
\frac{d \bm{J}}{dt}=-2\pi \int_{R_{\rm in}}^{R_{\rm out}} \frac{\bm{\omega }_{\rm p} \times
\bm{L}}{R^3}R\, dR,
\label{torque}
\end{equation}
where the integration is done over the surface of the disc.
\cite{MPT07} solve equation~(\ref{main}) to find the disc profile and
then use equation~(\ref{torque}) to find the timescale for alignment
of the black hole with the binary orbit. For $\beta=0$ this is
\begin{equation}
t_{\rm align}(0)= \frac{1}{\sqrt{2} \pi \Sigma}\left(\frac{acM_1}{\nu_{2}G}\right)^{\frac{1}{2}}
\end{equation}
\citep{SF}, where $\Sigma$ and $\nu_2$ must be evaluated at the warp
radius. The warp radius is found by balancing the Lense-Thirring
precession term with the viscous term associated with the second
viscosity in equation~(\ref{main}) to find
\begin{equation}
R_{\rm warp}=2\frac{\omega_{\rm p}}{\nu_2(R_{\rm warp})}
\end{equation}
\citep{SF,MPT07}.  In the next section we consider the size of this
radius for GRO J1655--40.

We can eliminate $\Sigma$ using the steady state accretion rate,
\begin{equation}
\dot M =3\pi\Sigma \nu_{1},
\end{equation}
to find
\begin{equation}
t_{\rm align}(0)= \frac{3 \nu_1}{\sqrt{2} \dot M}\left(\frac{acM_1}{\nu_{2}G}\right)^{\frac{1}{2}}.
\end{equation}
For $\beta \ne 0$ the alignment timescale is
\begin{equation}
t_{\rm align}(\beta)=t_{\rm align}(0) \tau,
\end{equation}
where
\begin{equation}
\tau=\frac{(1+\beta)^{-\frac{\beta}{1+\beta}}}{\sqrt{2}}
\frac{\Gamma \left(\frac{1}{2(1+\beta)}\right)}{\Gamma \left(\frac{1+2\beta}{2(1+\beta)}\right)\cos \left(\frac{\pi}{4(1+\beta)}\right)}
\label{tt0}
\end{equation}
\citep{MPT07} and $\nu_1$, $\nu_2$ and $\Sigma$ must be evaluated at
the warp radius, $R_{\rm warp}$.  The alignment of the system happens
exponentially. The time for the system to get very close to alignment
is actually several alignment timescales \citep{MPT07}.  In deriving
the timescale we assume that we can neglect the non-linear term
$\bm{l}.\partial^2\bm{l}/\partial R^2=-|\partial \bm{l}/\partial
R|^2$.  In the following section we calculate the alignment timescale
of GRO J1655--40.

\section{Black Hole Inclination}

We know that the inclination of the black hole to the orbital plane is
in the range $15^\circ<\eta<165^\circ$. We let $\eta$ be the current
inclination of the black hole to the outer disc. The outer disc is in
the binary orbital plane.  We let $\eta_0$ be the inclination when the
system first started transferring mass. If $\eta_0<\pi/2$ the system
started closer to alignment whereas if $\eta_0>\pi/2$ the system
started closer to counter-alignment than alignment. The system is
always aligning so $\eta<\eta_0$ even if it is counter-aligned. We use
the work of \cite{MPT07} to find that
\begin{equation}
\sin \eta = \sin \eta_0 \, e^{-t_{\rm accrete}/t_{\rm align}},
\label{less}
\end{equation}
where $t_{\rm accrete}$ is the time for which the system has been
steadily transferring mass, if $\eta_0<\pi/2$ and the system started
nearer to alignment than counter-alignment. However, if the system
started closer to counter-alignment then
\begin{equation}
\sin \eta = \sin \eta_0 \, e^{t_{\rm accrete}/t_{\rm align}}
\label{more}
\end{equation}
and $\eta_0>\pi/2$.  We estimate $t_{\rm accrete}$ from binary star
models in Section~\ref{bin} below.

\subsection{Co-rotating Disc and Black Hole}

\begin{figure}
  \epsfxsize=8.4cm 
\epsfbox{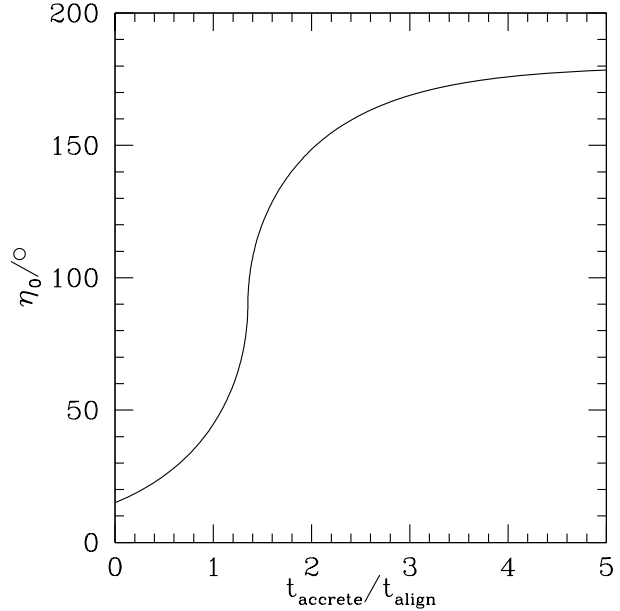}
\caption[]
{The initial inclination $\eta_0$ of the black hole relative to the binary
  orbit against the time that the disc has been accreting in a steady
  state. The disc has $\eta=15^\circ$ now.}
\label{fig:eta0}
\end{figure}

If the system started closer to alignment than counter-alignment so
that $\eta_0<\pi/2$ then we find the time that the accretion must last
for, by inverting equation~(\ref{less}), to be
\begin{equation}
t_{\rm accrete}=t_{\rm align}\log \left(\frac{\sin \eta_0}{\sin \eta}\right).
\end{equation}
However if the disc was initially counter-rotating so that
$\eta_0>\pi/2$ and now $\eta<\pi/2$ then we must find the time that it
takes to move from $\eta_0$ to $90^\circ$ using equation~(\ref{more})
and then add on the time it takes to move from $90^\circ$ to
$15^\circ$ using equation~(\ref{less}). Then we find
\begin{equation}
t_{\rm accrete}=-t_{\rm align}\log \left(\sin \eta_0\sin \eta\right).
\label{t2}
\end{equation}
As an example, in Fig.~\ref{fig:eta0} we see how the initial
inclination of the disc varies with the length of time it has been
steadily accreting for a disc which currently is at an inclination of
$15^\circ$.

\subsection{Counter-rotating Disc and Black Hole}

\begin{figure}
\epsfxsize=8.4cm 
\epsfbox{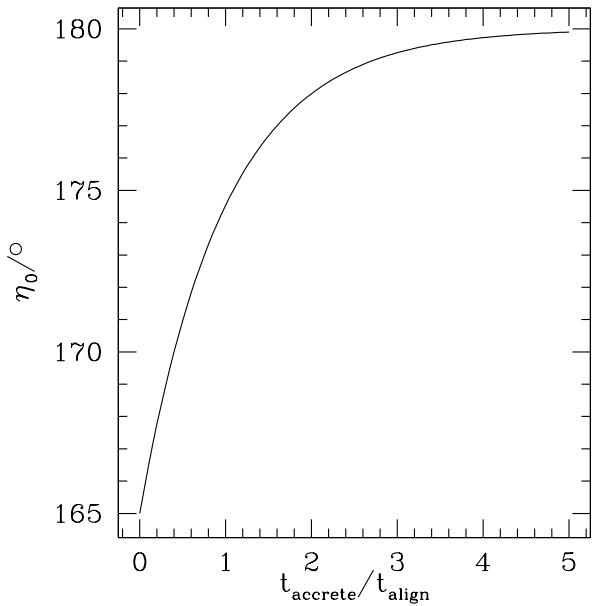}
\caption[]
{The initial inclination $\eta_0$ of the black hole relative to the binary
  orbit against the time that the disc has been accreting in a steady
  state. The disc has $\eta=165^\circ$ now.}
\label{fig:counter}
\end{figure}

If the disc is now closer to counter-alignment than alignment so that
$\eta>\pi/2$ then, using equation~(\ref{more}), we find the initial
inclination of the system as a function of the time for which it has
been accreting to be
\begin{equation}
t_{\rm accrete}=t_{\rm align}\log \left(\frac{\sin \eta}{\sin \eta_0}\right).
\label{t3}
\end{equation}
In Fig.~\ref{fig:counter} we see how the initial inclination of the
disc varies with the length of time it has been steadily accreting for
for a disc which is now at an inclination of $165^\circ$.

As illustrations we examine further these two extreme cases. If the
inclination of the black hole relative to the binary orbit is random
then the probability that it formed at $15^\circ$ or less is only
0.017.  Similarly the probability that it formed at $165^\circ$ or
more is only 0.017.  It is therefore likely that it has undergone some
but not excessive alignment and we would expect $t_{\rm
  accrete}\approx t_{\rm align}$.  We note that a completely
counter-aligned black hole is in an unstable equilibrium so it is
possible that we now observe the hole counter-aligned at an
inclination of $165^\circ$.  However this would have required both the
unlikely event of forming very close to counteralignment and the
accretion timescale to be small compared with the alignment timescale.
It is therefore more likely that the hole is now only $15^\circ$ from
alignment.  We cannot tell how misaligned it started but it may well
have been counter-aligned in the past.

\section{Warped Disc Model of GRO J1655--40}

In this section we find the viscosities, the warp radius and the
alignment timescale of GRO J1655--40.  The period of the system is
$2.62\,\rm d$ \citep{G01}.  By Kepler's law the separation is $R_{\rm
  B}=1.14\times 10^{12}\,\rm cm$ for a total mass of $8.7\,$M$_\odot$.
\cite{AK01} estimate that the dimensionless spin of the black hole
$a=0.2-0.67$, while \cite{WSO} suggest that it could be as high as
$0.92$.

\begin{figure}
\epsfxsize=8.4cm 
\epsfbox{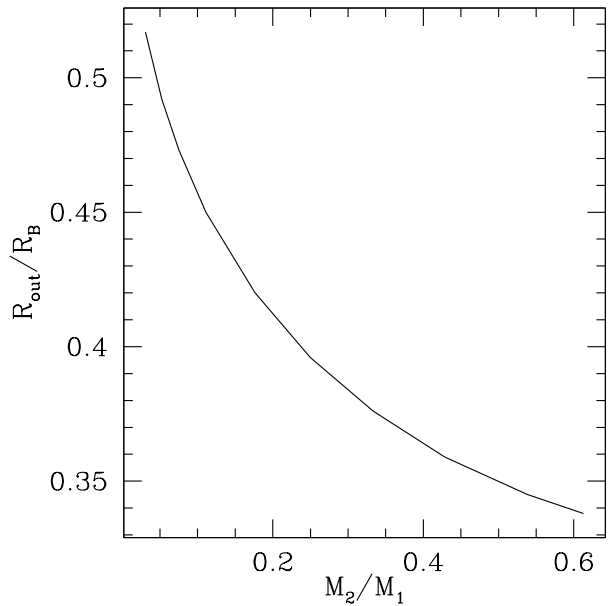}
\caption[]
{The truncation radius of the disc as a function of the ratio of the
  masses of the two stars.}
\label{fig:trunc}
\end{figure}

We take the outer truncation radius of the disc, $R_{\rm out}$, to be
the largest possible stable orbit of a test particle in this binary
system. We use parameters derived by \cite{P77} for the three body
system. In Fig.~\ref{fig:trunc} we plot the truncation radius against
the mass ratio of the two stars. GRO J 1655--40 has mass ratio
$M_2/M_1=2.4/6.3=0.38$ which corresponds to a truncation radius of
$R_{\rm out}=0.37\,R_{\rm B}=4.22\times 10^{11}\,$cm which is the
outer edge of our disc. This outer edge is used later to find
timescales in the disc.

\subsection{Viscosities}
\label{visct}

We use the $\alpha$-prescription to find the viscosity of the disc
\citep{SS73}.  In bright X-ray binaries the dominant radiative opacity
is bound free.  Using the steady state mass accretion rate, $\dot M
=3\pi \nu_1 \Sigma$, \cite{WP99} find the viscosity to be
\begin{align}
  \nu_j = &\ 6.40\times 10^{15}
  \alpha_j^{\frac{4}{5}}\left(\frac{M_1}{\rm
      M_\odot}\right)^{-\frac{1}{4}}\left(\frac{\dot M}{\rm
      10^{-8}\,M_\odot yr^{-1}}\right)^{\frac{3}{10}} \cr &\times
  \left(\frac{R}{10^{11}\rm cm}\right)^{\frac{3}{4}} \,\rm cm^2 g^{-1},
\label{visc}
\end{align}
where $j=1,2$, $\alpha_j$ is the usual $\alpha$-prescription for
viscosity $\nu=\alpha c_s H$ and $\dot M$ is the accretion rate on to
the black hole.  We take $\alpha_1=0.2$ and $\alpha_2=2$ \citep{LP07}.
This is equivalent to equations~(\ref{visc1}) and~(\ref{visc2}) with
$R_0=10^{11}\,\rm cm$ and
\begin{align}
\nu_{10}= &\ 1.11\times 10^{15} \left(\frac{\alpha_1}{0.2}\right)^{\frac{4}{5}}
 \left(\frac{M_1}{\rm 6.3\, M_\odot}\right)^{-\frac{1}{4}} \cr
 & \times \left(\frac{\dot M}{10^{-8}\rm M_\odot yr^{-1}}\right)^{\frac{3}{10}}\,\rm cm^2 g^{-1}
\label{nu1}
\end{align}
and
\begin{align}
  \nu_{20}=7.03\times 10^{15}\left(\frac{
      \alpha_2}{2}\right)^{\frac{4}{5}} \left(\frac{M_1}{\rm 6.3\,
      M_\odot}\right)^{-\frac{1}{4}}\cr \left(\frac{\dot M}{10^{-8}\rm
      M_\odot yr^{-1}}\right)^{\frac{3}{10}}\,\rm cm^2 g^{-1}.
\label{nu2}
\end{align}

The viscous timescale associated with $\nu_1$ is the timescale on
which matter is moved through the disc. For GRO J1655--40 this is
\begin{align}
  t_{\nu_1} = &\ \frac{R^2}{\nu_1}=1.7
  \left(\frac{\alpha_1}{0.2}\right)^{-\frac{4}{5}}
  \left(\frac{M_1}{\rm 6.3\, M_\odot}\right)^{\frac{1}{4}} \cr &
  \left(\frac{\dot M}{10^{-8}\rm M_\odot
      yr^{-1}}\right)^{-\frac{3}{10}}\left( \frac{R}{R_{\rm
        out}}\right)^{5/4}\,\rm yr.
\end{align}
This is a relatively short timescale and so the disc reaches steady
state quickly.

\subsection{Warp Radius}

We find the warp radius to be
\begin{equation}
R_{\rm warp}=\frac{2\omega_{\rm p}}{\nu_2(R_{\rm warp})}
= \frac{\omega_{\rm p}}{\nu_{20}(R_{\rm warp}/{\rm cm})^{\frac{3}{4}}}
\end{equation}
\citep{SF,MPT07} so that
\begin{align}
  R_{\rm warp} = &\ 4.33\times 10^8
  \left(\frac{a}{0.5}\right)^{\frac{4}{7}} \left(
    \frac{\alpha_2}{2}\right)^{-\frac{16}{35}} \left(\frac{M_1}{\rm
      6.3\, M_\odot}\right)^{\frac{9}{7}} \cr & \left(\frac{\dot
      M}{10^{-8}\rm M_\odot yr^{-1}}\right)^{-\frac{6}{35}}\,\rm cm.
\end{align}
We note that this is not strongly dependent on $\dot M$ and so the
outbursts of the system do not affect it greatly. We find the
viscosities and surface density at the warp radius so that we can find
the alignment timescale of the black hole with the disc.

\subsection{Timescale of Alignment}

In Section~\ref{gen} we derived a formula for the timescale on which
the black hole aligns with the disc. This assumes that the angular
momentum of the disc is much greater than that of the black hole. In
GRO J1655--40 the outer parts of the disc are locked to the binary
orbital plane by tidal torques. It is therefore the angular momentum
of the binary orbit, which is much greater than that of the black
hole, that matters because the disc is constantly fed with matter at
its outer edge..

\begin{figure}
\epsfxsize=8.4cm 
\epsfbox{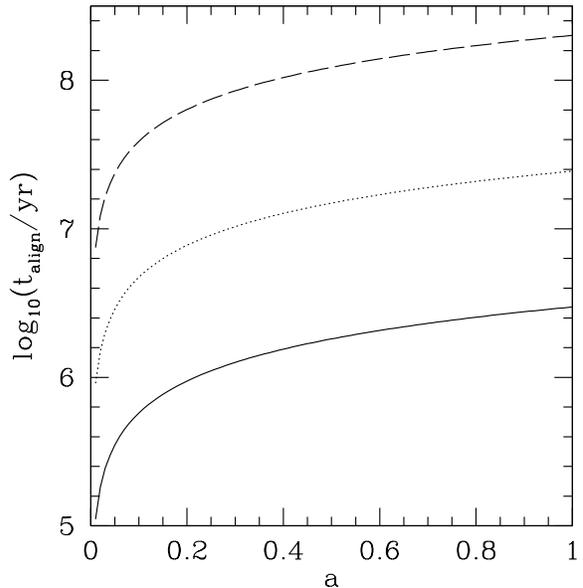}
\caption[]
{The timescale of the alignment of the black hole with the disc
  against $a$ for $\dot M=10^{-7}$ (solid line),~$10^{-8}$ (dotted
  line) and~$10^{-9}\,\rm M_\odot\, yr^{-1}$ (dashed line) }
\label{groa}
\end{figure}

\begin{figure}
\epsfxsize=8.4cm 
\epsfbox{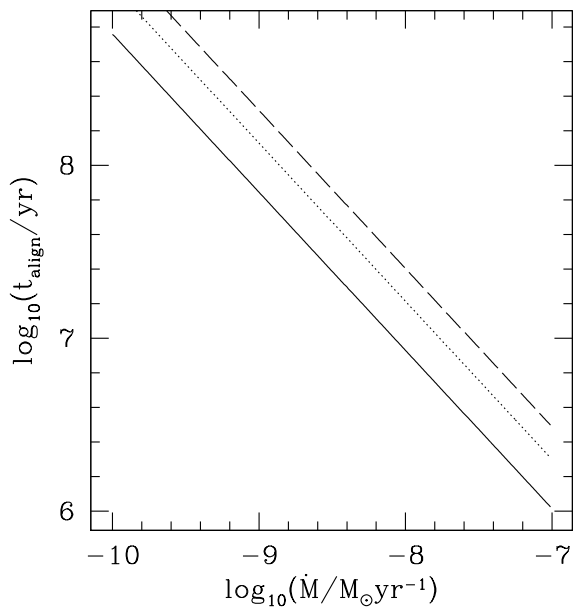}
\caption[]
{The timescale of the alignment of the black hole with the disc
  against $\dot M$ for $a=0.2$ (solid line),~$0.5$ (dotted line)
  and~$0.92$ (dashed line).  }
\label{gromd}
\end{figure}

We can now evaluate the timescale of alignment because we know the
viscosities and the warp radius at which we evaluate them.  For a
steady state disc we have $\beta=\frac{3}{4}$ (from
equations~(\ref{visc1}) and~(\ref{visc})) and so we find
\begin{align}
t_{\rm align}(3/4) &=  t_{\rm align}(0) \left(\frac{4}{7}\right)^{\frac{3}{7}}
\frac{\Gamma(2/7)}{\sqrt{2}\Gamma (5/7)\cos \left(\frac{\pi}{7}\right)}\cr
&= 1.5233 \, t_{\rm align}(0)
\label{factor}
\end{align}
and
\begin{align}
  t_{\rm align}(0)= &\ 9.75\times 10^{6}
  \left(\frac{a}{0.5}\right)^{\frac{5}{7}}
  \left(\frac{\alpha_1}{0.2}\right)^{\frac{4}{5}}
  \left(\frac{\alpha_2}{2}\right)^{-\frac{4}{7}} \cr &\ 
  \times\left(\frac{M_1}{\rm 6.3\, M_\odot}\right)^{\frac{6}{7}}
  \left(\frac{\dot M}{10^{-8}\rm M_\odot
      yr^{-1}}\right)^{-\frac{32}{35}} \,\rm yr
\end{align}
so that
\begin{align}
  t_{\rm align}(3/4)= &\ 1.49\times 10^{7}
  \left(\frac{a}{0.5}\right)^{\frac{5}{7}}
  \left(\frac{\alpha_1}{0.2}\right)^{\frac{4}{5}}
  \left(\frac{\alpha_2}{2}\right)^{-\frac{4}{7}}\cr & \times
  \left(\frac{M_1}{\rm 6.3\, M_\odot}\right)^{\frac{6}{7}}
  \left(\frac{\dot M}{10^{-8}\rm M_\odot
      yr^{-1}}\right)^{-\frac{32}{35}} \,\rm yr.
\label{talign}
\end{align}
In Fig.~\ref{groa} we plot the timescale for alignment against the
spin of the black hole, $a$, for a range of accretion rates. In
Fig.~\ref{gromd} we plot the timescale for alignment of the black hole
against the accretion rate for a range of spins. In the next section
we consider the time for which the mass transfer can take place to see
whether the black hole should be aligned or not.

\section{Binary Evolution models}
\label{bin}

For our equilibrium warped disc model to be valid, the viscous
timescale in the disc (about $1.7\,\rm yr$ as found in
section~\ref{visct}) must be much less than the time for which
material has been accreting through it.  We note that there has been
time since the 1994 outburst for the disc to regain equilibrium.
Outbursts have occurred more frequently in recent years but the system
generally returns to quiescence for sufficiently long periods for the
disc to regain equilibrium.  Because we still see the jet misaligned,
the timescale for alignment must be long enough relative to the
accretion timescale.  We therefore investigate the evolutionary state
of the system to determine both the expected mass-transfer rate and
the time since mass transfer began.

Early attempts to make detailed evolution models of GRO~J1655--40 met
with the problem that stars at the secondary's position in the
Hertzsprung-Russell diagram ought to be in the Hertzsprung gap and so
evolving from the end of their main sequence to the base of their
giant branch on a thermal timescale \citep{KKRF}.  As a consequence
the mass-transfer rate should be too high for the standard soft X-ray
transient explanation of the outbursts for which
\citet{vanparadijs1996} found that the average accretion rate on to
the black hole in GRO~J1655--40 should be less than $1.26\times
10^{-10}\,\rm M_\odot\,\rm yr^{-1}$.  Though we show that we can
easily make models in which the secondary star is still on the main
sequence, we still find that such a low mass-transfer rate is
inconsistent with any Roche-lobe filling system.

\citet{G01} have carefully observed the secondary star in GRO~J1655--40
during quiescence.  Their observations confirm that the secondary star
is filling its Roche lobe and are consistent with somewhat lower
masses for both the secondary star and the black hole.  Indeed we find
that we must put the secondary star right at the lower end of their
estimated mass range ($M_2 = 2.4\pm 0.4\,\rm M_\odot$) to fit its
position in the Hertzsprung-Russell diagram at all well.  The observed
mass function \citep{shahbaz1999} is
\begin{equation}
f(M_2) =\frac {(M_1\sin i)^3}{ (M_1 + M_2)^2} = 2.73\pm 0.09\,\rm M_\odot,
\end{equation}
where $M_1$ is the mass of the black hole and $i=70.2^\circ\pm
1.9^\circ$ \citep{G01} is the inclination of the binary orbit. In our
models we choose the current secondary mass $M_2$ and determine the
current mass of the black hole $M_1$ from the mass function and hence
the total mass $M_1 + M_2$ of the system.  We choose the initial
secondary mass $M_{2\rm i}$, the mass just before mass transfer
begins. Assuming conservation of mass and angular momentum during mass
transfer we have 
\begin{equation}
P(M_1M_2)^3 = \rm const
\end{equation}
so we can calculate the initial period $P_{\rm i}$ needed to give the
current period of $2.62168\pm 0.00014\,$d \citep{hooft1998}.
\citet{G01} measure a radius for the secondary star of $5.0\pm
0.3\,R_\odot$ which we combine with their chosen effective temperature
of $6{,}336\,$K to give a luminosity $L$ in the range $1.427 <
\log_{10} (L/L_\odot) < 1.667$.

We use the latest version of the Cambridge {STARS} code to construct
detailed evolutionary models of the secondary star.  The code was
originally written by \citet{eggleton1971,eggleton1972,eggleton1973}.
The equation of state, which includes molecular hydrogen, pressure
ionization and coulomb interactions, is discussed by \citet{pols1995}.
The initial composition is taken to be uniform with a hydrogen
abundance $X = 0.7$, helium $Y = 0.28$ and metals $Z = 0.02$ with the
meteoritic mixture determined by \citet{anders1989}.  Hydrogen burning
is allowed by the pp chain and the CNO cycles.  Reaction rates
are taken from \citet{caughlan1988}.  Opacity tables are those
calculated by \citet{iglesias1996} and \citet{alexander1994}.

\citet{foellmi2006} argue the source must be much nearer than typical
estimates of $3.2\,$kpc. However, Bailyn (private communication)
suggests that this is not possible. \citet{foellmi2006} assumed that
the secondary is a normal F6 IV star and so has the radius of an
isolated F6 IV star.  As we see below it is very easy to fit the
secondary with a somewhat more massive and luminous but Roche-lobe
filling star.

\begin{figure}
\epsfxsize=8.4cm
\epsfbox{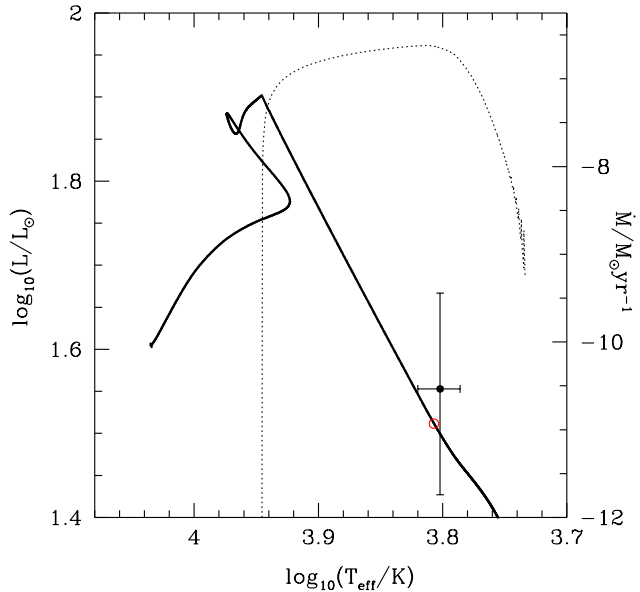}
\caption[]{The evolutionary track in the Hertzsprung-Russell diagram of
  a $2.5\,\rm M_\odot$ star in a binary and which is forced to
  transfer mass to its companion when it fills its Roche lobe which
  occurs while crossing the Hertzsprung gap.  The initial period and
  primary mass are chosen so that the period and masses fit those of
  GRO~J1655--40 when the secondary reaches $2\,\rm M_\odot$.  The
  dotted line and right-hand scale give the mass-transfer rate as a
  function of temperature.  The point with error bars is the observed
  secondary in GRO~J1655--40 and the open circle is the model with
  $M_2 = 2\,\rm M_\odot$ at the observed period.
\label{hr2p5}}
\end{figure}

\begin{figure}
\epsfxsize=8.4cm
\epsfbox{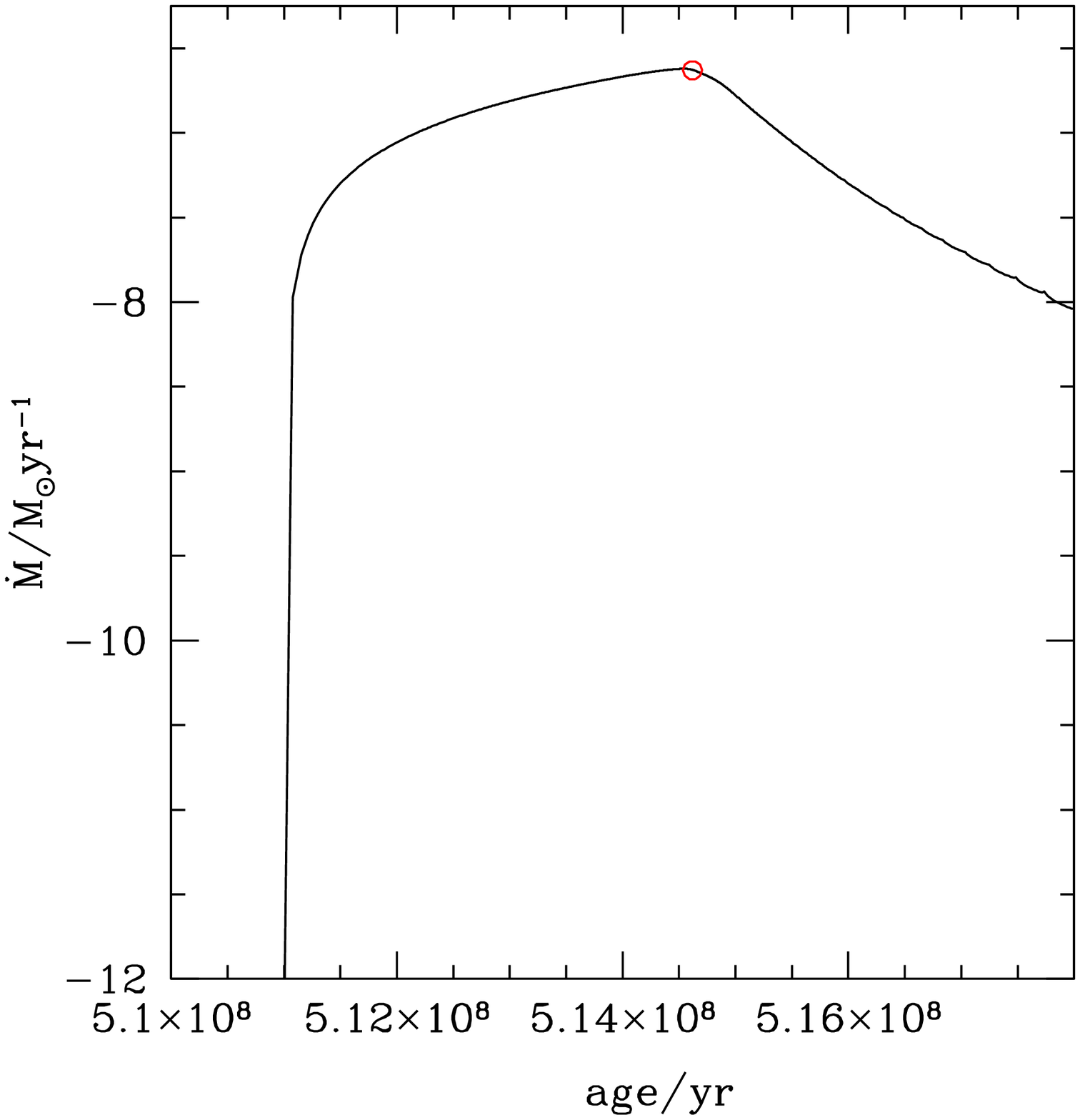}
\caption[]{The mass-transfer rate as a function of time for the system
illustrated in Fig.~\ref{hr2p5}.
\label{time2p5}}
\end{figure}

\begin{figure}
\epsfxsize=8.4cm
\epsfbox{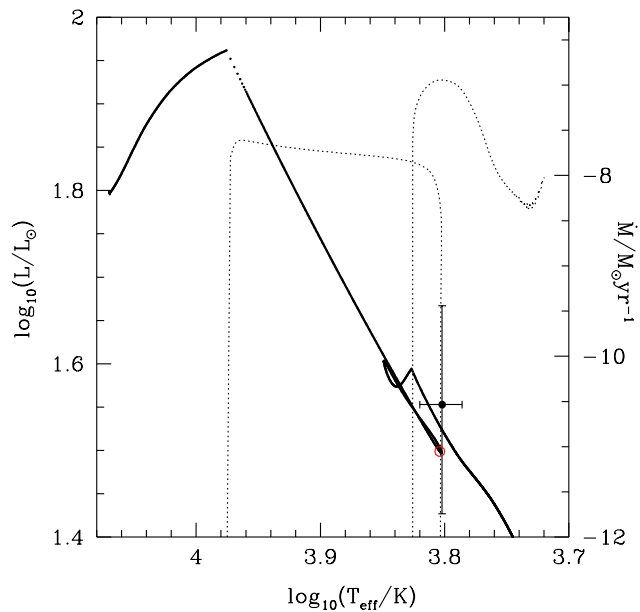}
\caption[]{As Fig.~\ref{hr2p5} but with a secondary star initially of
$2.8\,\rm M_\odot$.  This time mass transfer begins on the main sequence
when it is more gentle and lasts longer.
\label{hr2p8}}
\end{figure}

\begin{figure}
\epsfxsize=8.4cm
\epsfbox{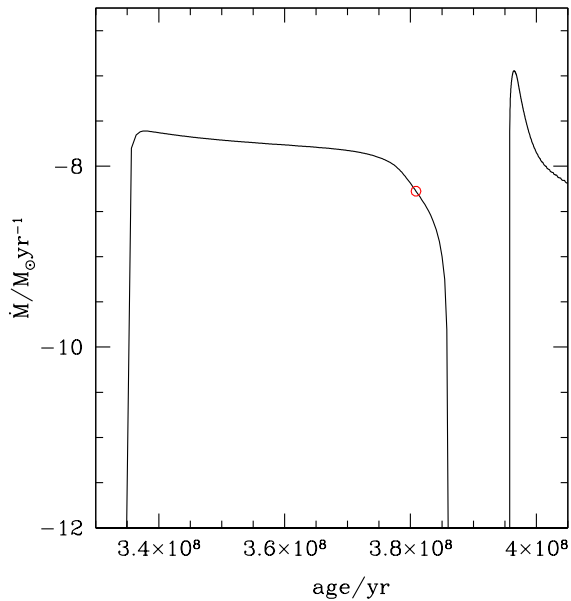}
\caption[]{The mass-transfer rate as a function of time for the system
illustrated in Fig.~\ref{hr2p8}.  
\label{time2p8}}
\end{figure}

After experimenting with a variety of initial conditions we find that,
to fit its position in the Hertzsprung-Russell diagram, the current
mass of the secondary star must be at the low end of the predicted
range, between about $2.0\,\rm M_\odot$ and $2.1\,\rm M_\odot$ and
that the mass of the black hole also lies at the low end of its
predicted range, between $5.88\,\rm M_\odot$ and $5.98\,\rm M_\odot$.
We illustrate two typical models in which the secondary is in
different evolutionary states.  

The first (Figs~\ref{hr2p5} and~\ref{time2p5}) has a secondary star,
of initial mass $2.5\,\rm M_\odot$, and initial period $1.648\,$d. It
fills its Roche lobe while evolving across the Hertzsprung gap.
Evolution during this phase is on a thermal timescale so that the mass
transfer rate is relatively high, $2.35\times 10^{-7}\,\rm M_\odot\,
yr^{-1}$, and lasts about $3\times 10^{6}\,$yr with $\dot M >
10^{-7}\,\rm M_\odot\,yr^{-1}$.  It fits the observations when the
secondary's mass has fallen to $2\,\rm M_\odot$ and the black hole
mass is $5.88\,\rm M_\odot$.  A total of $\Delta M=0.5\,\rm M_\odot$
of material has passed through the disc.  Similar results are found
for lower masses but both the accretion rate and the time for which
the black hole has been accreting fall.

The second (Figs~\ref{hr2p8} and~\ref{time2p8}) has a secondary star,
of initial mass $2.8\,\rm M_\odot$, and initial period $1.481\,$d.
This fills its Roche lobe while still evolving on the main sequence.
Evolution at this phase is on a nuclear timescale so that the mass
transfer rate is somewhat lower, $5.30\times 10^{-9}\,\rm M_\odot\,
yr^{-1}$, and lasts for about $5\times 10^{7}\,$yr.  Once again this
fits the observations when $M_2 = 2\,\rm M_\odot$ and $M_1 = 5.88\,\rm
M_\odot$.  In this time $\Delta M=0.8\,\rm M_\odot$ of material has
passed through the disc.  Secondary stars of initially higher mass
transfer material at a higher rate but not for much longer.  We also
experimented with including convective overshooting
\citep[c.f.][]{regos1998} to prolong the main-sequence lifetime but
found very little difference in the models that fit at these masses.

\section{Discussion}

We have successfully modelled GRO J1655--40 with mass transfer by
Roche-lobe overflow. We need the masses of the black hole and its
companion to be at the low end of their observed ranges. We find the
average mass-transfer rate and the length of time that it lasts for
different binary systems.

We have described two typical, but not unique, models in the previous
section. If the companion is crossing the Hertzsprung gap a typical
mass-transfer rate is $2.35\times 10^{-7}\,\rm M_\odot\, yr^{-1}$. For
this mass-transfer rate we predict an alignment timescale,
equation~(\ref{talign}), of $t_{\rm align}=7.8\times 10^5\,\rm yr$.
The accretion rate lasts for about $3\times 10^{6}\,$yr which is about
$3.8\,t_{\rm align}$.  For a secondary star still on the main sequence
a typical mass-transfer rate is $5.30\times 10^{-9}\,\rm M_\odot\,
yr^{-1}$. The alignment timescale for this lower rate is $t_{\rm
  align}=2.5\times 10^7\,\rm yr$. This accretion rate lasts for about
$5\times 10^{7}\,$yr which is about $2\,t_{\rm align}$. We note that
the alignment timescales also depend on the spin of the black hole
which is not known. In these estimates we have taken $a=0.5$.

In calculating the timescales we have assumed conservative mass
transfer and a constant accretion rate.  Given that mass is expected
to accumulate in the disc until an outburst the actual mass transfer
might be expected to take place in bursts that last for less than the
viscous timescale of the disc.  During the intervening quiescence,
when our equilibrium model can be applied, the mass-transfer rate
would be lower and the alignment timescale correspondingly longer.
However, a few tenths of a solar mass of material must pass through
the disc. The angular momentum of this material is much less than that
of the black hole.   Because $t_{\rm align} \propto \dot M^{-32/35}$,
$t_{\rm accrete}/t_{\rm align} \propto \Delta M\, \dot M^{-3/35}$
depends on $\dot M$ so changes in mass-transfer rate do not affect the
expected alignment very much.

We can also entertain the possibility that the average mass-transfer
rate might be low enough to be consistent with the
\citet{vanparadijs1996} SXT model.  In this case the alignment
timescale would be much longer.  We do not have a consistent
evolutionary model in this case but cannot expect the mass transfer to
last longer than the main-sequence lifetime of a $2\,\rm M_\odot$
star, about $8\times 10^8\,$yr.  We do not however expect that such a
low mass-transfer rate is possible from a Roche-lobe filling secondary
star other than for a short time at the onset and the end of mass
transfer.  It is unlikely that we are catching the system in either of
these brief periods and it is more likely that the condition for
transience needs to be reconsidered.

Alternatively the mass transfer might be highly non-conservative,
though there is no observational evidence for this.  In this case the
alignment timescale would be much longer while the evolutionary
timescales would be unchanged.  Thus the models we have discussed
would lead to the fastest alignment and this is just slow enough for
us to expect to find the black hole still misaligned with the binary
orbit.

We see that, whether the secondary fills its Roche lobe on the main
sequence or while evolving across the Hertsprung gap, the accretion
rate lasts for a few alignment timescales and so the fact that we see
the black hole still misaligned with the orbital plane now is
consistent.  We are unable to distinguish whether the black hole
is $15^\circ$ from alignment or counteralignment.  Given that the
misalignment is likely to be created by a strong kick at the time of
the supernova explosion we cannot expect either case to be more likely
than the other at the start of the accretion.  

For constant mass transfer our model predicts exponential decrease of
the misalignment.  We find that if $\eta = 15^\circ$ now and if mass
transfer has been going on for $2 t_{\rm align}$ then the misalignment
would have initially been $\eta_0=148^\circ$. If the black hole has
been accreting for $3.8\,t_{\rm align}$ then the system would have
initially been at $\eta_0=175^\circ$.  However these are only
indicative models and it is relatively easy to find others in which
the accretion has been going on for less time, but still similar to
the alignment timescale.  We conclude that the system may or may not
have been counter-aligned after its supernova kick but that it is most
likely to be close to alignment rather than counteralignment now
because otherwise the black hole would have had to form very close to
$\eta_0 = 180^\circ$ and this is rather unlikely.

\section{Conclusions}

We have found that the secondary star is most likely evolving either
at the end of the main sequence or across the Hertzsprung gap. The
mass-transfer rate is typically fifty times higher in the latter case
but we find that, in both cases, the lifetime of the mass transfer
state is at most a few times the alignment timescale.  The fact that
the black hole has not yet aligned with the orbital plane is therefore
consistent with either model. The system may or may not have been
counter-aligned after its supernova kick but we conclude that it is
now more likely to be close to alignment rather than counteralignment.

\section*{Acknowledgements}

We thank Tom Maccarone for useful discussions of the alignment
timescale of the system and Charles Bailyn for discussion of the
latest observations of the system.  CAT thanks Churchill College for a
Fellowship.

\label{lastpage}
\end{document}